\documentclass[12pt]{iopart}
\usepackage{graphicx}
\usepackage{amssymb}
\begin{document}

\title[Chi-squared Test for Binned, Gaussian Samples]{Chi-squared Test for Binned, Gaussian Samples}

\author{Nicholas R. Hutzler}

\address{Division of Physics, Mathematics, and Astronomy \\ California Institute of Technology \\ Pasadena, CA 91125}
\ead{hutzler@caltech.edu}
\vspace{10pt}
\begin{indented}
\item[]June 2019
\end{indented}

\begin{abstract}
We examine the $\chi^2$ test for binned, Gaussian samples, including effects due to the fact that the experimentally available sample standard deviation and the unavailable true standard deviation have different statistical properties.  For data formed by binning Gaussian samples with bin size $n$, we find that the expected value and standard deviation of the reduced $\chi^2$ statistic is
\medskip
\begin{equation}  \frac{n-1}{n-3}\pm \frac{n-1}{n-3}\sqrt{\frac{n-2}{n-5}}\sqrt{\frac{2}{N-1}},  \end{equation}
\medskip
where $N$ is the total number of binned values.  This is strictly larger in both mean and standard deviation than the value of $1\pm (2/(N-1))^{1/2}$ reported in standard treatments, which ignore the distinction between true and sample standard deviation.
\end{abstract}

%
% Uncomment for keywords
%\vspace{2pc}
%\noindent{\it Keywords}: XXXXXX, YYYYYYYY, ZZZZZZZZZ
%
% Uncomment for Submitted to journal title message
%\submitto{\JPA}
%
% Uncomment if a separate title page is required
%\maketitle
% 
% For two-column output uncomment the next line and choose [10pt] rather than [12pt] in the \documentclass declaration
%\ioptwocol
%

\newcommand{\nbin}{n}
\newcommand{\var}{\textrm{Var}}
\newcommand{\cov}{\textrm{Cov}}
\newcommand{\EE}{\textrm{E}}
\newcommand{\std}{\textrm{Std}}
\newcommand{\normdist}{\mathcal{N}}
\newcommand{\chisq}{\chi^2}
\newcommand{\samplechisq}{\tilde{\chi}^2}
\newcommand{\samplechi}{\tilde{\chi}}
\newcommand{\invchisq}{\textrm{Inv-}\chi^2}
\newcommand{\redchisq}{\chi_{red}^2}
\newcommand{\sampleredchisq}{\tilde{\chi}_{red}^2}
\newcommand{\truestd}{\sigma}
\newcommand{\truemean}{\mu}
\newcommand{\Ny}{N}
\newcommand{\be}{\begin{equation}}
\newcommand{\ee}{\end{equation}}
\newcommand{\bea}{\begin{eqnarray}}
\newcommand{\eea}{\end{eqnarray}}

\section{Introduction}

Precision measurements of physical quantities typically require a very large number of individual measurements of the same quantity often taken under varying conditions, such as drifting signal-to-noise or many experimental configurations with different signal sizes.  For this reason, as well as for simplification of data analysis and reduction of computational requirements, the data are typically binned together such that measurements in the same bin were taken within a time during which the conditions were similar.  In order to check whether the binning is susceptible to the varying conditions, as well as to search for unknown sources of noise, a $\chisq$ test~\cite{Press2007,Bevington2003,Taylor1996} is commonly used.  Regardless of whether or not it is an ideal choice of statistic for this case, it is fairly intuitive as a measure of whether the assigned error bars are correctly capturing the statistics of the data.  However, some of the simplifying assumptions used to construct the standard $\chisq$ can give results with a significant bias for large data sets.  We discuss why the standard treatment underestimates both the mean and variance of the $\chisq$ statistic, and then determine the appropriate correction factors.

\section{Chi-squared test for binned, Gaussian samples}

Consider a quantity $N_x\gg 1$ of measurements $x_i$ without any assigned uncertainties.  Say that the measurements are normally distributed with constant, true mean $\mu$ that is not known to the experimenter.  We shall not assume that the data has a constant variance.  Let us gather these data sequentially into groups $G_j$ with $\nbin$ consecutive points each.  Now compute the usual sample mean, standard deviation, and standard error of each group of points:
\be y_j = \frac{1}{\nbin}\sum_{x_i \in G_j}x_i, \quad
s_j = \sqrt{\frac{1}{\nbin-1}\sum_{x_i \in G_j} (x_i-y_j)^2}, \quad
s_{yj} = \frac{1}{\sqrt{\nbin}}s_j.\ee

%\begin{figure}[htbp]
	%\centering
		%\includegraphics[width=0.45\textwidth]{binning_schematic.pdf}
	%\caption{Illustration of the binning procedure for $n=10$.  We will gather $10$ of the $x_i$ points (sequentially) and compute the mean and standard error for each bin, $ y_j\pm s_{yj}$.  The true mean $\mu=0$ is fixed, while the true (unknown) standard deviation $\sigma$ is slowly varying.}
	%\label{fig:binning_schematic}
%\end{figure}

We have now binned our data into a smaller set of $\Ny = N_x/\nbin\gg 1$ mean values $y_j$ and uncertainties $s_{yj}$.  As a check to see whether the assigned uncertainties are correctly capturing the statistical fluctuations of the data we can perform a $\chi^2$ test as outlined in many standard texts~\cite{Press2007,Bevington2003,Taylor1996}.  We will test the hypothesis that the $y_j$ are normally distributed about a constant $\bar{y}$ (though this approach is easily extended to models with more degrees of freedom), and that the uncertainties correctly describe the statistical fluctuations of the data about the mean.  The reduced-$\chi^2$ value of the data set is
\be \redchisq = \frac{1}{\Ny-1}\sum_{j=1}^{\Ny} \left(\frac{y_j - \bar{y}}{\sigma_{yj}}\right)^2 \equiv \frac{1}{\Ny-1}\sum_{j=1}^\Ny \chi_j^2,\label{eq:redChiSq}\ee
where $\bar{y} = {(\sum_j y_j/s_{yj}^2)}/{(\sum_j 1/s_{yj}^2)}$ is the weighted mean of the $y$ data, and $\sigma_{yj}$ is the true (unknown) standard deviation of the points $\{x_i\in G_j\}$, which need not be constant over different values of $j$.  If the fluctuations in the data are Gaussian in nature, and correctly accounted for by the uncertainties, then we have the usual result
\be
\EE[\redchisq] = 1, \quad 
\std[\redchisq] =\sqrt{\frac{2}{\Ny-1}}.
\ee
However, the experimenter does not know the true standard deviation, and therefore actually computes the statistic
\be \sampleredchisq = \frac{1}{\Ny-1}\sum_{j=1}^{\Ny} \left(\frac{y_j - \bar{y}}{s_{yj}}\right)^2 \equiv  \frac{1}{\Ny-1}\sum_{j=1}^{\Ny} \samplechisq_j, \ee
using $s_{yj}$ as an estimator for $\sigma_{yj}$.  We wish to find the statistical properties of this quantity, which we shall find differ from $\redchisq$.  Intuitively, the sample standard deviation is computed from a finite number of measurements and therefore has some uncertainty associated with it, and that uncertainty should be propagated through when examining the $\sampleredchisq$ statistic.  This is a well-known effect when estimating parameters from finite data sets and has been previously explored in a number of contexts, for example Poisson distributions, counting experiments, weighted means, and histogram fitting~\cite{Baker1984,Jading1996,Hammersley1997,Mighell1999,Hauschild2001,Zhang2006,Gagunashvili2010}.

More specifically, while $\chi_j\sim\normdist(0,1)$ is normally distributed, $\samplechi_j$ is not:
\be \samplechi_j \equiv \left(\frac{y_j - \bar{y}}{s_{yj}}\right) \approx \left(\frac{y_j - \mu}{s_{yj}}\right) \sim t(\nbin-1), \ee 
the $t$-distribution with $n-1$ degrees of freedom, which has larger tails for finite $\nbin$ than a normal distribution.  Notice that we are treating $\bar{y} = \mu$ as a constant, which is valid in the limit $\Ny\gg 1$, though for smaller $\Ny$ the statistical properties of the weighted mean cannot be ignored~\cite{Zhang2006,Hutzler2014Thesis,Cochran1937,Graybill1959,Bockenhoff1998,Hartung2008}.  In particular, the weighted mean also has correction factors due to the difference between true and sample standard deviation, and has a non-trivial variance, both of which will impact the $\sampleredchisq$ statistic.  A good discussion of these complexities can be found in reference~\cite{Hartung2008}.

The square of $\samplechi_j$ is therefore distributed as $\samplechisq_j\sim F(1,\nbin-1)$, the $F$-distribution with $(1,\nbin-1)$ degrees of freedom, which has
\be \EE[F(1,\nbin-1)] = \frac{\nbin-1}{\nbin-3} ,\quad \var[F(1,\nbin-1)] = 2\left(\frac{\nbin-1}{\nbin-3}\right)^2\frac{\nbin-2}{\nbin-5}.  \ee
This is as opposed to the $\chisq_j$ statistic, which has (appropriately) a $\chisq$ distribution.  $\sampleredchisq$ is therefore distributed as a sum of $F$-distributions, which is complicated~\cite{Morrison1971}.  However, the expectation value and variance are straightforward to calculate,
\bea
\EE[\sampleredchisq]
& = & \frac{\Ny}{\Ny-1}\EE\left[\samplechisq_j\right] \nonumber\\
& = & \frac{\nbin-1}{\nbin-3} + \mathcal{O}\left(\Ny^{-1}\right), \label{eq:EE}\\
\var[\sampleredchisq] 
& = & \frac{\Ny}{(\Ny-1)^2}\var\left[\samplechisq_j\right] \nonumber\\
& = & \frac{2}{\Ny-1}\left(\frac{\nbin-1}{\nbin-3}\right)^2\frac{\nbin-2}{\nbin-5}  + \mathcal{O}\left(\Ny^{-2}\right)\label{eq:var}.
\eea

\begin{figure}[htbp]
	\centering
		\includegraphics[width=0.35\textwidth]{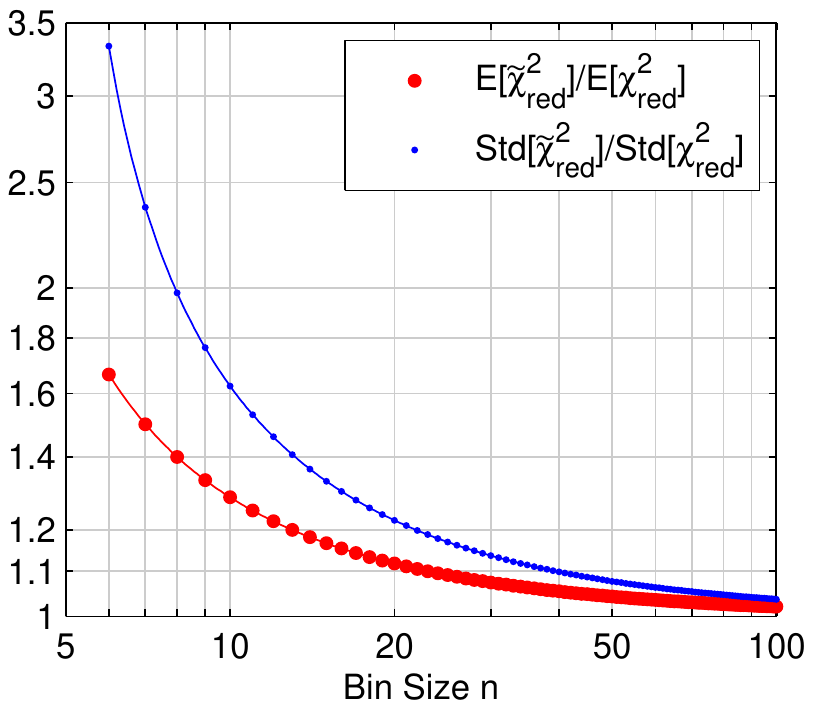}
	\caption{Correction factors to the mean and standard deviation of $\sampleredchisq$.}
	\label{fig:chi2_correction_factors}
	\vspace{-5mm}
\end{figure}

This implies that the mean and standard deviation of the $\sampleredchisq$ statistic are larger than those of the $\redchisq$ statistic by
\be
\frac{\EE[\sampleredchisq]}{\EE[\redchisq]} = \frac{\nbin-1}{\nbin-3}, \quad 
\frac{\std[\sampleredchisq]}{\std[\redchisq]} =\frac{\nbin-1}{\nbin-3}\sqrt{\frac{\nbin-2}{\nbin-5}},
\ee
up to further corrections of order $\mathcal{O}\left(\Ny^{-1}\right)$.  A plot of these correction factors is shown in Figure~\ref{fig:chi2_correction_factors}.  In the limit $\nbin\rightarrow\infty$ we recover the usual result, but for finite $\nbin$ we will always expect larger values for both mean and standard deviation.  We can also see that choosing $n\leq 5$ is not advisable, since the statistic will have a non-convergent variance.

\section{Conclusion}

In summary, we find that the standard $\chi^2$ statistic computed from binning finite data sets underestimates the mean and variance for binned Gaussian samples, and derive simple, closed expressions for the biases.  For very large data sets with finite bin sizes, such as those commonly found in precision physics measurements, these corrections can be significant and should not be neglected.

\emph{Acknowledgments.}  I would like to acknowledge helpful discussions with David Watson, and many helpful discussions with the ACME Collaboration, in particular David DeMille, John M. Doyle, and Brendon O'Leary.

\section*{Appendix: A simple example}

We can see how the ``usual'' chi-squared statistic gives an incorrect result by performing a simple numerical test on some simulated data.  Generate 1,000,000 points $x_i\sim\normdist(0,1)$, bin into groups of $\nbin=10$, and then compute means $y_j$, standard errors $\sigma_{yj},$ and the reduced chi-squared statistic $\sampleredchisq$ (as described in the main text) for the resulting 100,000 binned points.
\begin{quote}
\begin{verbatim}
Nx = 1000000 //Number of x values
nbin = 10 //Number of points to bin
for j = 1:(Nx/nbin) //Step over bins
    x = randn(1,nbin) //Generate nbin normally distributed points
    y(j) = mean(x) //Means
    sigmayi(j) = std(x)/sqrt(nbin) //Standard errors
end
ybar = sum(y./sigmayi.^2)/sum(1./sigmayi.^2) //Weighted mean
chi = (y-ybar)./sigmayi //chi
chi2 = sum(chi.^2) //chi^2
dof = length(y)-1 //Degrees of freedom
redchi2 = chi2/dof //Reduced chi^2
redchi2sigma = sqrt(2/dof) //``Usual'' uncertainty of chi^2
\end{verbatim}
\end{quote}
If we run this piece of code, we will find \texttt{redchi2 = 1.2868} and \texttt{redchi2sigma = 0.0045} (though of course the former will be different each time due to the random nature of the calculation.)  This value differs considerably from the na\"{i}ve expectation of $1\pm0.0045$ based on the usual treatment that ignores the difference between sample and true standard deviations, but is quite close to the expected value of $1.2857\pm0.0073$ from equations (\ref{eq:EE}) and (\ref{eq:var}).

\newpage

\bibliography{library}

\end{document}